\documentstyle[twoside,fleqn,espcrc2,epsfig,amsfonts]{article}

\newcommand{\AmS}{{\protect\the\textfont2
  A\kern-.1667em\lower.5ex\hbox{M}\kern-.125emS}}

\title{Study of the order of the phase 
	transition in pure U(1) gauge theory  with 
	Villain action}

\author{J. Cox\address{Center for Theoretical Physics, MIT, Cambridge
        MA, U.S.A.},
	T. Neuhaus\address{Helsinki Institute of Physics,
	 Finland}
        and 
        H. Pfeiffer\address{Institut f{\"u}r Theoretische Physik E,
        RWTH Aachen, Germany}}
       
\begin{document}

\begin{abstract}
We address the question of the order of the deconfinement phase
transition of four dimensional U(1) lattice gauge theory. 
Simulations of the $\mathbb{Z}$-gauge
theory dual to the Villain action on toroidal lattices up to lattice
sizes of $28^4$
give results consistent with both, a vanishing and a nonvanishing
discontinuity in the thermodynamic limit. A decision on the order
of the phase transition requires still larger lattice sizes.
\end{abstract}

\maketitle

\section{INTRODUCTION}
\vspace{-0.3cm}
U(1) lattice gauge theory with Villain action
\begin{equation}
\exp(-S_{\rm V})=\prod_{\rm P}\sum_{\rm m_{\rm P}\in{\mathbb{Z}}}\exp\left\{
         -\frac{\beta}{2}(\theta_{\rm P}-2\pi m_{\rm P})^2\right\}\nonumber
\label{vill}
\end{equation}
is, on the infinite lattice, equivalent~\cite{Pe78} to the 
${\mathbb{Z}}$ gauge theory
\begin{equation}
S_{\rm Z}=\kappa\sum_{\rm P}n_{\rm P}^2\qquad \left(\kappa=
	\frac{1}{2\beta}\right),
\label{zgauge}
\end{equation}
a gauge theory with an integer valued gauge field on the links.
In particular the order of the phase transition and universal 
quantities are the same.
We choose the ${ \mathbb{Z}}$ gauge theory for computational advantages and use
a toroidal lattice with periodic boundary conditions.
We expect\footnote{This is suggeted by a Fourier transform of $S_{\rm V}$.} 
the Villain action to have properties similar to those of the
extended action
\begin{equation}
  S_{\rm E}=-\sum_P(\beta\cos(\theta_P) + \gamma\cos(2\theta_P))
\label{ext}
\end{equation}
with $\gamma\approx -0.22$.\\

The order of the phase transition in this parameter region
is controversial.
Scaling of
gauge-balls~\cite{CoFr97b} suggests second order. Finite size scaling
of the specific heat
is consistent with first order~\cite{CaCr98a,CaCr98b} (torus and 
surface of 5D hypercube)  as well as with
second order~\cite{LaPe96,JeLa96} (homogenous sphere).
The volume dependence 
of the latent heat on finite lattices is consistent with 
first order as well as with second order.

For a more detailed discussion of our results we refer to~\cite{CoNe98}.
\vspace{-0.4cm}
\section{SIMULATION}
\vspace{-0.3cm}
\begin{table}[hbt]
\newlength{\digitwidth} \settowidth{\digitwidth}{\rm 0}
\catcode`?=\active \def?{\kern\digitwidth}
  \begin{center}
    \leavevmode
    \footnotesize
    \begin{tabular}{|c|l|l|r|r|r|r|}\hline
      $L^4$&$\kappa$&$\kappa_{\rm pc}(C_{\rm V})$&Msw.& $\tau_{\rm int}$& $\tau_{\rm flip}$\\
      \hline  
      \hline  
      $6^4$ &0.77955 &0.780350(18)&29&101  &   -         \\
      \hline  	       			   
      $8^4$ &0.77716 &0.778332(24)&10.5&233  &  -          \\
            &0.77822 &            &14.1&228  &  -          \\
      \hline  	       			   
     $10^4$ &0.77713 &0.777556(21)& 4.3&502  &  -         \\
            &0.77745 &&1.035&479  &       -    \\
      \hline  	       			   
     $12^4$ &0.77706 &0.777183(12)&2.05 &887  &     -      \\
            &0.77715 &&11.5&935  &       -    \\
      \hline  	       			   
     $14^4$ &0.77699 &0.776976(07)&9.9 &1697 &    9000    \\
      \hline  	       			   
     $16^4$ &0.77689 &0.776882(11)&7.8 &2852 &   12500    \\
      \hline  	       			   
     $18^4$ &0.77681 &0.776816(07)&4.42 &4384 &  20000     \\
            &0.77685 &            &1.525&5156 &  19000     \\
      \hline  	       			   
     $20^4$ &0.77677 &0.776786(06)&2.9 &8639 &     27000  \\
            &0.77680 &            &1.05&7084 &     31000  \\
      \hline  	       			   
     $22^4$ &0.77676 &0.776760(06)& 2.55 &12903&   43000    \\
            &0.77677 &            & 2.862&11383&   47000     \\
      \hline  	       			   
     $24^4$ &0.77674 &0.776742(05)& 2.0 &16360&   40000    \\
            &0.77675 &            & 2.1 &17807&   50000    \\
      \hline  	       			   
     $26^4$ &0.776745&0.776758(10)&1.42 &35151&  110000    \\
      \hline  	       			   
     $28^4$ &0.77672 &0.776726(07)&0.95 &28701&  120000    \\
      \hline  
    \end{tabular}
\caption{Lattice sizes, $\kappa$ values and positions of the specific
heat maxima. The statistics is given in $10^6$ sweeps. Also included
are the integrated autocorrelation time and an estimate of the flip
duration in sweeps.}
\label{tab:stat}
\end{center}
\vspace{-1.1cm}
\end{table}
A Metropolis algorithm is used for the update.
The statistics is summarized in Tab.~\ref{tab:stat}.
We perform Ferrenberg Swendsen multihistogram
reweighting~\cite{FeSw89}
 to the maximum of
the specific heat. Jackknife errors are calculated for all quantities.
The flip autocorrelation time $\tau_{\rm flip}$, which 
corresponds to
the mean life time of a pure phase configuration, is estimated
from the time evolution of the energy density.
$\tau_{\rm flip}$ is a more adequate measure for the number of
statistically independent measurements than $\tau_{\rm
int}$. $\tau_{\rm flip}$  is in our case about a factor 3-5
greater than $\tau_{\rm int}$.

\vspace{-0.4cm}
\section{SPECIFIC HEAT}
\begin{figure}[htb]
\begin{center}	
\epsfig{file=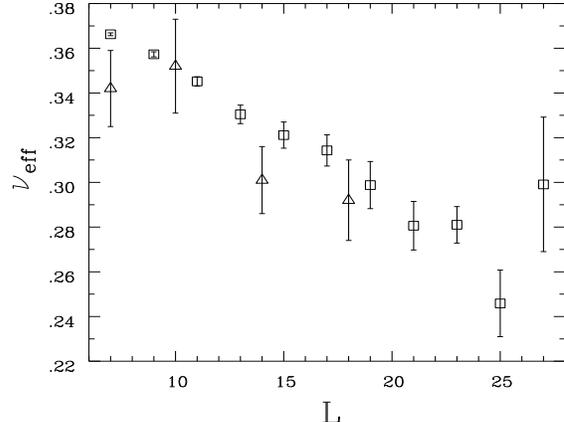,
width=6.0cm,height=7.5cm,angle=90,
bbllx=30,bblly=200,bburx=530,bbury=760}
\vspace{-2.0cm}
\caption{The effective exponent $\nu_{\rm eff}(L)$ for the Villain
action~(\ref{vill})~(squares) and the extended  
action~(\ref{ext})~(triangles) at
$\gamma=-0.2$. The values for the extended action are
taken from Fig.~15 in~\cite{CaCr98b}.}
\label{fig:nu}
\end{center}
\vspace{-1.2cm}
\end{figure}

\vspace{-0.3cm}
Finite size scaling theory for first order transitions (see e.g.~\cite{Bi87}) 
predicts the
maximum of the specific heat to diverge $\propto L^D$ which implies
$\nu=1/4$. The effective exponent $\nu_{\rm eff}(L)$, which is calculated from
the specific heat at two neighbouring $L$,  is
shown in Fig.~\ref{fig:nu}.  
We compare with $\nu_{\rm eff}$
obtained in~\cite{CaCr98b} at $\gamma=-0.2$ and find agreement almost within
error bars.\\
Clearly there is a tendency for $\nu_{\rm eff}(L)$ 
to decrease over the $L$ range
considered, but a stabilization on a certain value is not observed.

\vspace{-0.4cm}
\section{PROBABILITY DIS\-TRI\-BUTION}
\vspace{-0.3cm}
The probability distribution $p_L(e)$, where $e$ is the action divided
by the number of plaquettes, has for $L\geq 10$ two peaks
with a minimum in between them. It is not well described by the sum of
two Gaussians.
The minimum becomes more pronounced with increasing volume, but
the distance between the peaks decreases.
\begin{figure}[htb]
\begin{center}	
\vspace{-0.2cm}
\epsfig{file=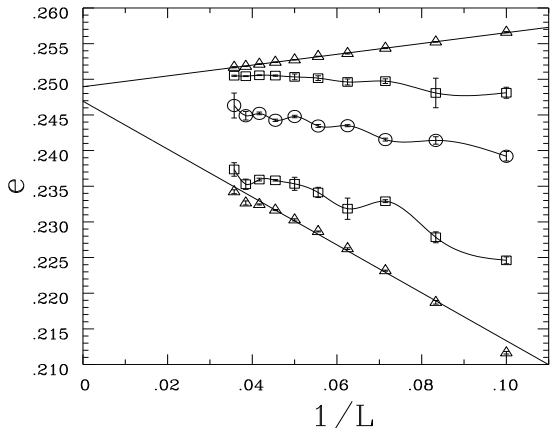,
width=5.5cm,height=9.5cm,angle=90,
bbllx=70,bblly=35,bburx=530,bbury=760}
\vspace{-1.4cm}
\caption{The positions of the maxima (squares), of the minimum
(circles) of the energy density probability distribution, and
the position on the flanks, where the probability distribution has
half the value of the respective maximum (triangles).}
\label{fig:pos}
\end{center}
\vspace{-1.3cm}
\end{figure}
The histograms are reweighted to the maximum of the specific heat. For a first
order transition this is asymptotically equivalent to equal weight of
the coexisting phases.
Then the positions in $e$ of the maxima, the minimum and where the
histogram is half of its respective maximum value on the
outer flanks of the distribution function are calculated.

A first order transition would obviously require the width of $p_L(e)$ to
extrapolate to a non-zero value.
It is consistent with the data that the histogram width shrinks
to zero approximately like $1/L$ for $L\longrightarrow\infty$
(Fig.~\ref{fig:pos}).
This behaviour would imply 
a second order phase transition in spite of the observation of
hysteresis and a double peaked histogram on finite lattices.

The position especially of the lower maximum (confinement phase) varies
irregularly. Presumably the discrete nature of the considered action
contributes to this effect.

The $L$ extrapolation of the upper energy 
branch's data (Coulomb phase) is less sensitive to the 
extrapolation ansatz than the lower energy branch's
data. This is related to the fact 
that the energy fluctuations within the confinement branch 
are much larger than in the Coulomb phase. In fact they have 
the magnitude of the possible gap itself.

Following~\cite{BiNe93}, we use
\begin{equation}
  F(L)=\frac{1}{2L^{d-1}}\log\left(\frac{P_L^{\rm max,1}P_L^{\rm
	max,2}}{(P_L^{\rm min})^2} \right)
\end{equation}
as a finite volume estimator for the interface tension.
$F(L)$ slowly decreases with the lattice size. Again a stabilization
to a constant value is not observed. It may shrink to zero
for $L\rightarrow\infty$.
A power law fit $F(L)\propto L^{-z}$ results in $z=0.3(1)$.

Also the ratio of the peak heights is not
stable in the considered $L$ range.
\vspace{-0.4cm}
\section{CONCLUSIONS}
\vspace{-0.3cm}
We have done a high statistics simulation of the ${\mathbb{Z}}$ gauge theory
using large lattices.
Nevertheless, it is not possible to give definite answers to the question of
the order of the phase transition from our simulations. We find that 
the typical indications of a first order transition --- hysteresis 
and histograms with double peaks --- persist up to $L=28$. On the other
hand, the width of the energy density histogram might shrink to
zero, which would imply second order. The $L$ dependent exponent $\nu_{\rm
eff}(L)$ decreases steadily but does not stabilize in the considered $L$
range. The limit $\lim_{L\rightarrow\infty}\nu_{\rm eff}(L)$ requires
larger lattices.  
The finite volume estimator for the interface tension
slowly decreases with $L$ and might vanish in the thermodynamic limit.
In addition, the flattening of
the probability distribution in the metastable phase region
typical for first order phase transitions has not
been observed in our simulations. Therefore a stabilization of the
finite volume interface tension estimator is not expected.
The ratio of the peak heights at the maximum of $C_V$ does not
stabilize in the considered $L$ range.

The simulations have been done on toroidal lattices as opposed to
spherical lattices with homogenously distributed
curvature~\cite{LaPe96,JeLa96}, where full consistency with
second order finite size scaling behaviour has been seen.
Also, we want to point out that a
mass determination on toroidal lattices away from the transition point
--- where finite size effects are under control ---, as has been done 
in~\cite{CoFr97b}, is feasible in spite of the complications
immediately at the transition point. Those results favor second order.
Our data for the effective $\nu_{\rm eff}(L)$ indicate that if the  
phase transition should actually turn out to be first order, 	 
the next larger lattice  sizes  may exhibit a stabilization of
$\nu_{\rm eff}(L)$ at the value 1/4. Such lattices could, in principle,
be simulated on massively parallel architectures.

The transition
might well be of first order, but even if that was a given fact larger 
lattices would be necessary for a better understanding of the peculiar finite 
size effects. 
In that case the similarity to second order would indicate
a critical point not far away in the coupling space.
In any case it will be interesting to pursue the
investigation of pure U(1) gauge theory on the lattice further.
\vspace{-0.4cm}
\section{ACKNOWLEDGEMENTS}
\vspace{-0.3cm}
This work is supported in part by funds provided by the
U.S. Department of Energy (D.O.E.) under cooperative research
agreement DE-FC02-94ER40818.
The computations have been performed on the Cray-T90 of HLRZ
Juelich. We thank I.~Campos, J.~Jers{\'a}k and U.-J. Wiese for
discussions.

\vspace{-0.2cm}
\bibliographystyle{wunsnot}   

\end{document}